\documentclass[aps,prl,twocolumn,amsmath,amssymb,longbibliography,superscriptaddress]{revtex4-2}
\pdfpageattr {/Group << /S /Transparency /I true /CS /DeviceRGB>>}
\usepackage{graphicx}
\usepackage{color}
\usepackage[hidelinks]{hyperref}
\usepackage{orcidlink}
\usepackage{amsmath}
\usepackage{amsthm}
\usepackage{physics}
\usepackage{amsfonts}
\usepackage{times,txfonts}
\usepackage{lipsum}

\hypersetup{colorlinks=true, linkcolor=blue, citecolor=blue, urlcolor=blue}

\begin{document}

\author{George Mihailescu\,\orcidlink{0000-0002-0048-9622}}
\email[]{george.mihailescu96@gmail.com}
\affiliation{School of Physics, University College Dublin, Belfield, Dublin 4, Ireland}
\affiliation{Centre for Quantum Engineering, Science, and Technology, University College Dublin, Dublin 4, Ireland}

\author{Karol Gietka\,\orcidlink{0000-0001-7700-3208}}
\email[]{karol.gietka@uibk.ac.at}

\affiliation{Institut f\"ur Theoretische Physik, Universit\"at Innsbruck, Technikerstra{\ss}e\,21a, A-6020 Innsbruck, Austria} 

\title{Mind the Gap: Anti-Critical Quantum Metrology}


\begin{abstract}
Critical quantum metrology exploits the dramatic growth of the quantum Fisher information near quantum phase transitions to enhance the precision of parameter estimation. This enhancement is commonly associated with a closing energy gap, which causes the characteristic timescales for adiabatic preparation or relaxation to diverge with increasing system size. As a consequence, the apparent growth of the quantum Fisher information largely reflects the increasing evolution time induced by critical slowing down rather than a genuine improvement in metrological performance, thereby limiting the practical usefulness of such protocols. Here we show that the relationship between energy gaps, quantum correlations, and achievable precision in interacting quantum systems can be far more subtle. In particular, quantum-enhanced sensitivity can also emerge when the energy gap increases, eliminating critical slowing down and enabling substantially faster relaxation dynamics. Although the corresponding quantum Fisher information may decrease due to the shorter evolution time, the resulting precision can nevertheless remain quantum-enhanced. Building on this insight, we introduce an anti-critical quantum metrology scheme in which quantum-enhanced precision arises while the energy gap grows. We illustrate this mechanism using the quantum Rabi model, thereby identifying a route to metrological advantage that avoids the slow dynamics associated with conventional criticality.
\end{abstract}
\date{\today}
\maketitle


\emph{Introduction.}---Quantum metrology seeks to exploit the laws of quantum mechanics to estimate physical parameters with the highest possible precision~\cite{QM2006Llyod,QM2011Lloyd}. Quantum correlations can be used to surpass the standard quantum limit of precision achievable with separable states, reaching in principle the Heisenberg limit~\cite{PhysRevLett.102.100401,PhysRevLett.105.180402,elusiveHL2012}. Despite tremendous experimental and theoretical progress, overcoming the standard quantum limit in macroscopic quantum systems remains largely elusive~\cite{PhysRevLett.110.181101,RevModPhys.89.035002,RevModPhys.90.035005,AdvancesNatphotson2018,photonicQM2020,6v93-whwq}. The reason is that strongly correlated many-body quantum states are difficult to prepare and are extremely fragile to noise and decoherence~\cite{noisyQM2011, PhysRevLett.111.120401}. This has motivated the search for new approaches, and over the past decade critical quantum metrology has emerged as a particularly promising direction~\cite{PhysRevA.78.042105,PhysRevLett.99.100603,PhysRevLett.99.095701,mihailescu2025criticalquantumsensingtutorial}. 

\begin{figure}[htb!]
    \centering
    \includegraphics[width=0.8\linewidth]{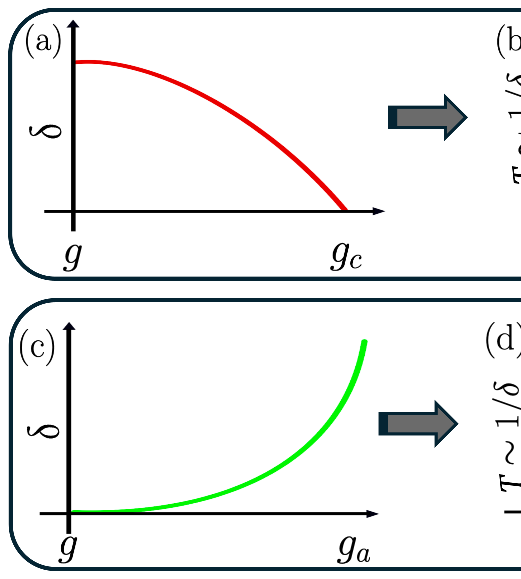}
    \caption{Schematic illustration of the energy gap \(\delta\) and associated relaxation timescales \(T \sim \delta^{-1}\) as a function of a control parameter \(g\) for a static system. (a) In critical quantum systems, tuning the system to its critical point, \(g \to g_c\), leads to a closing of the energy gap. (b) Correspondingly, the characteristic timescale grows. This results in a longer waiting time required to extract the signal, and an increase duration of each measurement cycle. (c) Gap opening as control parameter \(g\) is tuned. (d) This corresponds to a decrease in the characteristic timescale: allowing for fast measurement repetition cycles. In anti-critical metrology, one seeks to generate the same degree of non-classical correlations as critical quantum systems, but couple this with fast dynamics, achieved by tuning the system to the corresponding ``anti-critical'' point \(g \to g_a\).}
    \label{fig:comparisond}
\end{figure}

By tuning a system close to a quantum phase transition, the closing of the many-body energy gap and the associated build-up of quantum correlations lead to extreme sensitivity, as witnessed by the quantum Fisher information. In some situations, this sensitivity can be more robust than in conventional approaches to quantum metrology~\cite{PhysRevA.109.L050601, PhysRevLett.133.040801,PhysRevA.111.052621,cxvs-5pb1}. However, the same gap closing that boosts the quantum Fisher information is also related to critical slowing down: where the characteristic timescales for preparation, transformation, probing, and readout scale as $T \sim 1/\delta$ and diverge as the gap vanishes (see Fig.~\ref{fig:comparisond} for a schematic illustration). In many circumstances, this can be a severe limiting factor, especially in the presence of decoherence. Consequently, the enhanced sensitivity signaled by a large quantum Fisher information is often nullified by the prohibitively long evolution and preparation times~\cite{Hayes_2018,PhysRevX.8.021022}. This trade-off is especially consequential in the frequentist setting, where saturating the Cramér--Rao bound requires many independent repetitions, increasing the time cost of each trial.

In this work, we challenge the prevailing intuition that closing energy gaps is required for quantum-enhanced metrology in interacting systems~\cite{MONTENEGRO20251}. While a vanishing gap guarantees a large quantum Fisher information, quantum enhancement can also emerge when the gap remains finite—or even increases. This counterintuitive regime arises when the opening of the energy gap, largely overlooked in previous studies, plays the central role. Unlike critical metrology, where enhanced sensitivity comes with critical slowing down, here the system develops strong quantum correlations while remaining fast enough to equilibrate and enable measurements on short timescales. This defines a form of anti-critical metrology that circumvents critical slowing down without sacrificing correlation-induced advantages, although the quantum Fisher information itself may be reduced due to the shorter evolution time. For this reason, we introduce an operationally relevant figure of merit for ground state quantum metrology, the gap-regularized quantum Fisher information, which also accounts for the total time needed to prepare and interrogate the probe state~\cite{PRXQuantum.6.020351}. 

In essence, anti-critical metrology relies on the simultaneous buildup of quantum correlations and gap opening. We illustrate this mechanism using the quantum Rabi model~\cite{PhysRevLett.115.180404}, which has been extensively studied in the context of critical quantum metrology~\cite{Gietka2021adiabaticcritical,PhysRevLett.126.010502,PRXQuantum.3.010354,PhysRevLett.124.120504,chen2025globalized}. Our results re-frame the notion that gap closing is a prerequisite for quantum enhancement and identify fast many-body dynamics as a complementary resource for quantum sensing. Although we focus on the quantum Rabi model for concreteness, the underlying mechanism—interaction-induced squeezing combined with gap engineering—is general and not restricted to integrable or few-mode systems. Its applicability is, however, limited to systems that allow controllable gap opening~\cite{SM}.

\emph{Equilibrium metrology and the role of the gap.}---Quantum metrology concerns estimating an unknown parameter, here denoted by $\omega$, and the fundamental precision limits imposed by quantum mechanics. The quantum Cramér--Rao bound~\cite{PhysRevLett.72.3439,paris2009,RevModPhys.90.035005}
\begin{equation}
    \Delta^{2}\omega \,\geq\, \frac{1}{\mu\,\mathcal{I}_{\omega}}
\end{equation}
relates the minimum mean-square error of any locally unbiased estimator to the quantum Fisher information $\mathcal{I}_{\omega}$ and the number of independent realizations $\mu\gg1$. In Hamiltonian parameter estimation, the sensitivity of an equilibrium state such as the ground state $\lvert\psi_{0}\rangle$ is determined by
\begin{equation}
    \label{eq:qfi_gs}
    \mathcal{I}_{\omega} \,=\, 4\big[\langle \partial_{\omega}\psi_{0} \vert \partial_{\omega}\psi_{0}\rangle \,-\, \vert \langle \partial_{\omega}\psi_{0} \vert \psi_{0}\rangle \vert^{2}\big]\,.
\end{equation}
The physical origin of large metric responses becomes transparent in the spectral representation,
\begin{equation}
    \label{eq:qfi_spectral}
    \mathcal{I}_{\omega} \,=\, 4\sum_{n\neq 0}\,\frac{\big\vert \langle \psi_{n}(\omega)\vert \partial_{\omega}\hat{H} \vert \psi_{0}(\omega)\rangle \big\vert^{2}}{\big[E_{n}(\omega)-E_{0}(\omega)\big]^{2}}\,,
\end{equation} 
where inverse powers of excitation gaps appear explicitly. 
Thus the quantum Fisher information is controlled by intrinsic response timescales set by the excitation spectrum; the slowest timescale scales as $T \sim \delta^{-1}=(E_1-E_0)^{-1}$ and diverges as the spectral gap $\delta$ closes~\cite{PhysRevX.8.021022}. Consistently, the equilibrium bound~\cite{PRXQuantum.6.030309} constrains
\begin{equation}
    \mathcal{I}_{\omega} \,\leq\, \frac{\Vert \partial_{\omega}\hat{H}\Vert^{2}}{\delta^{2}}\,,
\end{equation}
showing that divergences of the quantum Fisher information can be induced by a vanishing spectral gap without quantum enhancement related to quantum correlations, as exemplified in the End Matter.

To separate genuine quantum enhancement from amplification due solely to gap closing and the associated long timescales, we introduce the gap-regularized quantum Fisher information
\begin{equation}
    \label{eq:qfi_gap}
    \bar{\mathcal{I}}_{\omega} \,\equiv\, \delta^{2}\,\mathcal{I}_{\omega}\,,
\end{equation}
where $\delta$ is the relevant spectral gap. This quantity enables fair comparisons across systems with different gap structures and intrinsic timescales. Gapped, or gap-opening, interacting systems can exhibit substantial sensitivity from collective effects and correlations; while their quantum Fisher information may decrease as the gap grows, their bounded response times can still make them practically advantageous.


\emph{Quantum Rabi Model.}---To illustrate the idea of anti-critical metrology, we consider the quantum Rabi model~\cite{Xie_2017}, which is one of the simplest models exhibiting a finite-component quantum phase transition~\cite{PhysRevLett.115.180404,QRM2021QPT}. Originally introduced to describe light–matter interactions in cavity quantum electrodynamics, the quantum Rabi model focuses on a single mode of the electromagnetic field interacting with a two-level system. However, it is a universal model for a harmonic oscillator coupled to a two-level system and can be realized in a wide variety of physical platforms, including trapped ions~\cite{QRM2015trappedions,PhysRevX.8.021027}, circuit quantum electrodynamics~\cite{cQED2017QRM}, and ultracold atoms~\cite{PhysRevResearch.4.043074}. The Hamiltonian of the quantum Rabi model reads ($\hbar \equiv 1$)
\begin{align}
    \hat{H} = \omega \hat{a}^\dagger \hat{a} + \frac{\Omega}{2} \hat{\sigma}_z + \frac{g}{2} (\hat{a} + \hat{a}^\dagger)  \hat{\sigma}_x,
\end{align}
where $\omega$ is the frequency of the harmonic oscillator, $\Omega$ is the energy splitting of the two-level system, $g$ is the light–matter coupling strength, $\hat{a}$ and $\hat{a}^\dagger$ are the bosonic annihilation and creation operators, and $\hat{\sigma}_x$ and $\hat{\sigma}_z$ are Pauli matrices acting on the two-level system.

In this work, we focus on the finite-component phase transition that emerges in the regime where the two-level splitting is much larger than the oscillator frequency, $\Omega \gg \omega$~\cite{PhysRevLett.115.180404}. In this limit, the hierarchy of energy scales effectively suppresses spin flips and renders the spin degree of freedom increasingly rigid. As a result, the ratio $\Omega/\omega$ plays a role analogous to an effective system size and defines the thermodynamic limit of the model. Taking $\Omega/\omega \to \infty$ therefore constitutes the thermodynamic limit in which nonanalytic behavior can arise, despite the presence of only a single two-level component coupled to an oscillator. To leading order, one finds
\begin{align}
\hat{H}_{\mathrm{eff}}
= \omega\, \hat{a}^\dagger \hat{a}
+ \frac{\Omega}{2}\, \hat{\sigma}_z
+ \frac{g^{\,2}}{4\Omega}\, (\hat{a} + \hat{a}^\dagger)^2 \hat{\sigma}_z
+ \mathcal{O}\!\left(\frac{\omega}{\Omega}\right),
\end{align}
which already captures the essential features of a quantum phase transition like closing energy gap and the built-up of correlations.

For the low energy spin down sector (effectively replacing $\hat \sigma_z$ with $-1$) we obtain up to constant terms
\begin{align}
    \hat{H}\approx \omega \hat{a}^\dagger \hat{a}   - \frac{g^2}{4\Omega} \left(\hat{a} + \hat{a}^\dagger\right)^2,
\end{align}
which corresponds to a single-mode squeezing Hamiltonian whose ground state is a squeezed vacuum
\begin{align}
    |\psi\rangle = \exp\!\left[\tfrac{\xi_-}{2}( \hat a^2 +  \hat a^{\dagger 2})\right] |0\rangle,
\end{align}
where $\xi_- \equiv -\tfrac{1}{4}\log(1-g^2/g_c^2)$ is the squeezing parameter. Equivalently, the squeezing Hamiltonian can be expressed in the diagonal form, leading to an opening harmonic oscillator
\begin{align}\label{eq:effectiveH}
    \hat{H}\approx \omega \sqrt{1-\frac{g^2}{\omega \Omega}} \hat{c}^\dagger \hat{c},
\end{align}
where $\hat c = \hat a\cosh \xi_- - \hat a^\dagger \sinh \xi_-$ is the annihilation operator associated with the new mode with frequency $\omega \sqrt{1-g^2/\omega\Omega}$.

The finite-component quantum phase transition occurs when the coefficient in front of $(\hat{a} + \hat{a}^\dagger)^2$ reaches the critical value at  $g =g_c \equiv\sqrt{\omega \Omega}$ at which the effective oscillator frequency $\omega \sqrt{1-{g^2}/g_c^2}$ vanishes, leading to a closing of the energy gap. While conventional critical metrology would operate in the vicinity of this critical point~\cite{PhysRevLett.124.120504,PhysRevResearch.4.043074,PhysRevLett.130.090802}, where the quantum Fisher information diverges, the anti-critical approach deliberately avoids the critical point and the associated critical slowing down. Instead, we engineer parameter regimes in which the correlations as well as the estimation precision increase while the effective energy gap becomes larger. This allows us to retain the metrological enhancement of correlated states without the detrimental scaling of timescales (see Fig.~\ref{fig:fig1} for a schematic illustration). In the context of the quantum Rabi model (see Supplemental Material~\cite{SM} for alternative models), one can achieve such an opening by considering the high-energy sector and replacing $\hat \sigma_z \to +1$. The resulting effective Hamiltonian reads
\begin{align}\label{eq:antiH}
     \hat{H}\approx \omega \hat{a}^\dagger \hat{a}   + \frac{g^2}{4\Omega} (\hat{a} + \hat{a}^\dagger)^2,
\end{align}
which is again a squeezing Hamiltonian. Its ground state is a squeezed vacuum,  
\begin{align}
    |\psi\rangle = \exp\!\left[\tfrac{\xi_+}{2}( \hat a^2 +  \hat a^{\dagger 2})\right] |0\rangle,
\end{align}
but with a different squeezing parameter $\xi_+ \equiv -\tfrac{1}{4} \log(1+g^2/g_c^2)$. This corresponds to an effective harmonic oscillator
\begin{align}\label{eq:effant}
    \hat H = \omega\sqrt{1+g^2/g_c^2}\, \hat c^\dagger \hat c = \omega e^{-2\xi_+}\, \hat c^\dagger \hat c ,
\end{align}
with a frequency that grows as the coupling strength $g$ is increased. Thus squeezing is generated while the energy gap increases rather than closes (see Fig.~\ref{fig:fig1}c). In the following, we revise the critical metrology approach and introduce the anti-critical metrology protocol.

\begin{figure}
    \centering
    \includegraphics[width=0.9\linewidth]{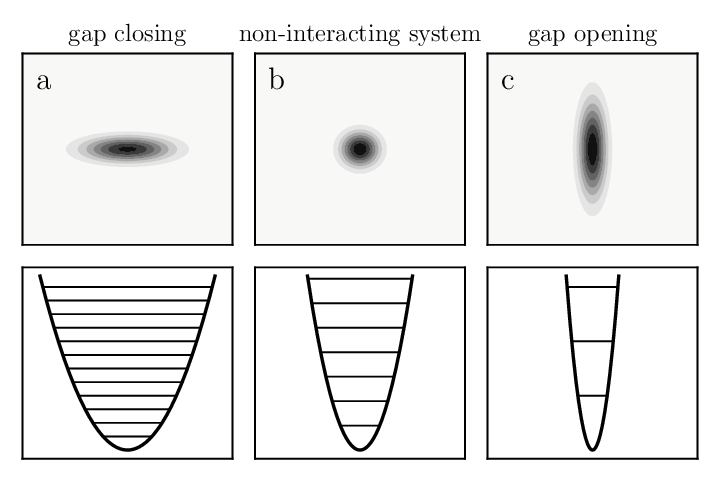}
    \caption{Schematic illustration of critical and anti-critical metrology using the effective description of the quantum Rabi model.
The bottom panels show the associated effective potentials, while the top panels depict the corresponding ground states in phase space. Starting from the noninteracting spectrum and vacuum state (b), interactions can either close the energy gap—realizing critical metrology (a)—or open it—realizing anti-critical metrology (c). In both cases the final state is squeezed, but the characteristic timescales differ dramatically. In critical metrology, the vanishing gap leads to long timescales and a large quantum Fisher information, whereas in anti-critical metrology a growing gap shortens the timescales and reduces the quantum Fisher information. This behavior reflects the fact that, in many interacting quantum systems, the energy gap—and therefore the relevant dynamical timescales—scale with system size.}
    \label{fig:fig1}
\end{figure}

\emph{Critical Metrology and Anti-Critical Metrology.}---In the quantum Rabi model, deep in the $\Omega \gg \omega$ limit, the ground state in both the critical and anti-critical regimes reduces to a squeezed vacuum
\begin{align}
    |\psi\rangle = \exp\!\left[\tfrac{\xi_\pm}{2}( \hat a^2 +  \hat a^{\dagger 2})\right] |0\rangle,
\end{align}
with squeezing parameter $\xi_\pm \equiv -\tfrac{1}{4}\log(1\pm g^2/g_c^2)$, where sub-indices $\pm$ correspond to the anti-critical and critical regimes, respectively. 
The squeezing parameter is directly related to the frequency of the effective harmonic oscillator~\cite{mirkhalaf2025frequencyshiftsreflectionground},
\begin{equation}
\omega \sqrt{1\pm g^2/g_c^2} = \omega e^{-2\xi_\pm}.
\end{equation}
As the interaction strength $g$ increases, the squeezing grows in both regimes, which is well known to enhance metrological precision. 
However, the spectral gap behaves oppositely: in the critical regime it closes as the critical point is approached, leading to diverging timescales~\cite{PhysRevLett.95.105701,Dziarmaga01112010,RevModPhys.83.863,PhysRevA.105.042620} (see Fig.~\ref{fig:fig1}a), whereas in the anti-critical regime the gap opens with increasing $g$, resulting in progressively shorter timescales.

Evaluating the quantum Fisher information for this squeezed ground states yields  
\begin{align}
\begin{split}
    \mathcal{I}_\omega =\frac{\Delta^2 \hat a^\dagger \hat a }{\omega^2 \left(1 \pm g^2/g_c^2\right)}\sim \langle \hat a^\dagger \hat a \rangle^2 T_\pm^2,
 \end{split}
\end{align}
where $T_\pm^{-2} \sim  \omega^2(1\pm g^2/g_c^2)$ is related to the energy gap inverse squared. In both cases the quantum Fisher information exhibits quadratic (Heisenberg) scaling with the number of excitations $\langle \hat a^\dagger \hat a\rangle$ and time $T_\pm$ which is related to the energy gap. The core difference between the two approaches lies in the relation between the excitation content and the time-scales related to the energy gap.

To make this explicit, consider first the mean excitation number of the squeezed vacuum in the critical approach,
\begin{align}
\langle \hat a^\dagger \hat a\rangle = \sinh^2\xi_-
\approx \frac{1}{4\sqrt{1-g^2/g_c^2}},
\end{align}
where the approximation holds close to the critical point $g\approx g_c$. The characteristic timescale is constrained by the closing of the energy gap~\cite{PhysRevLett.95.105701,Dziarmaga01112010,RevModPhys.83.863,PhysRevX.8.021022,gyhm2025fundamentalscalinglimitcritical,abiuso2025informationtheoreticproofplanckianbound}. For instance, during an adiabatic sweep from $g/g_c \approx 1-\epsilon$ (with $\epsilon\ll 1$) toward the critical point $g/g_c \approx 1$, the minimal evolution time scales as
\begin{align}\label{eq:Tcrit}
T_- \sim \frac{1}{\omega\sqrt{1-g^2/g_c^2}}
\sim \frac{4\langle \hat a^\dagger \hat a\rangle}{\omega}.
\end{align}
Thus, both the excitation number and the evolution time inherit the same critical dependence on the parameters of the system. Substituting these relations into the quantum Fisher information gives~\cite{PhysRevLett.124.120504}
\begin{align}\label{eq:QFIcrit}
    \mathcal{I}_\omega = \frac{g^4/g_c^4 }{8 \omega ^2 \left(1-{g^2}/{g_c^2 }\right)^2}\sim \langle \hat a^\dagger \hat a\rangle^2\, T_-^2,
\end{align}
showing that the apparent divergence of the quantum Fisher information at $g\sim g_c$ is entirely accounted for by Heisenberg scaling with respect to the excitation number and with respect to the characteristic time. 

In the anti-critical metrology approach on the other hand, the mean excitation number grows with coupling as
\begin{align}
    \langle \hat a^\dagger \hat a\rangle = \sinh^2 \xi_+ \approx \tfrac{1}{4}\, g/g_c,
\end{align}
valid for $g \gg g_c$, while the characteristic time decreases as
\begin{align}\label{eq:anticT}
    T_+ \sim \frac{1}{\omega g/g_c} \sim \frac{1}{\omega \langle \hat a^\dagger \hat a\rangle}.
\end{align}
As a consequence, the quantum Fisher information becomes constant
\begin{align}
    \mathcal{I_\omega} = \frac{1}{8 \omega^2} \sim  \langle \hat a^\dagger \hat a\rangle^2\, T_+^2.
\end{align}

\begin{figure}[ht!]
    \centering
    \includegraphics[width=0.9\linewidth]{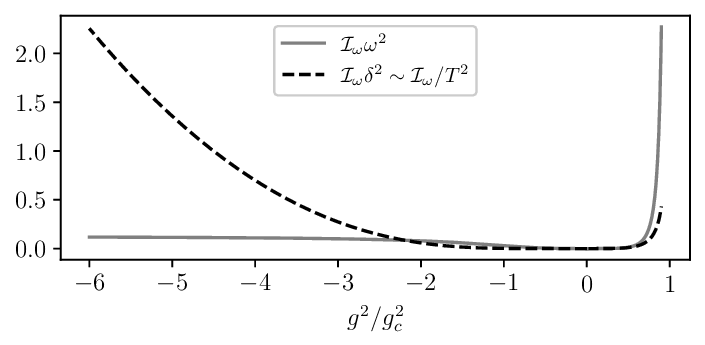}
    \caption{Comparison between the quantum Fisher information (solid gray line) and the regularized quantum Fisher information (dashed black line).
While the quantum Fisher information diverges near the critical point ($g^{2}/g_{c}^{2}\approx1$), accounting for the time reveals that the sensitivity can be equally large away from criticality, where the required evolution time is much shorter due to the absence of critical slowing down. Exploiting the opening of the energy gap, and thus enabling an anti-critical metrology approach, may however require significantly stronger interactions. Simulations are performed using the effective description of the quantum Rabi model from Eq.~\eqref{eq:effectiveH} for $g^2/g_c^2 >0$, and from Eq.~\eqref{eq:antiH} for $g^2/g_c^2<0$.}
    \label{fig:comparison}
\end{figure}

Although both approaches exhibit the same dependence of the quantum Fisher information on the number of excitations and the evolution time, their physical origin differs substantially. In the critical case, the quantum Fisher information is effectively inflated by the long preparation times associated with a closing spectral gap, whereas in the anti-critical (gap-opening) regime it is reduced by the short response times enforced by a finite gap. Employing the gap-regularized quantum Fisher information reveals that, at fixed resources—such as a fixed number of excitations—the resulting precision arising from quantum correlations can be comparable in the two approaches. The anti-critical regime may nevertheless be advantageous in situations where rapid state preparation and short interrogation times are desirable, as it avoids the slow dynamics associated with criticality. See Fig.~\ref{fig:comparison} for a comparison between the critical and anti-critical approaches using both the quantum Fisher information and its energy-gap–regularized form.

\emph{Effective oscillator viewpoint.}---A complementary perspective is provided by the effective harmonic oscillator description [Eqs.~\eqref{eq:effectiveH} and~\eqref{eq:effant}] relevant in the ultra-strong coupling regime~\cite{RevModPhys.91.025005,USC2019nature,cxvs-5pb1}, where the environment can couple through the eigenstates of $\hat{c}$ rather than those of the bare mode $\hat{a}$~\cite{PhysRevA.84.043832,stefanini2025lindbladme,10.21468/SciPostPhysLectNotes.68}. In this picture, the parameter $\omega$ is estimated from the response of an oscillator with renormalized frequency $\omega_{\mathrm{eff}}(\omega) = \omega \sqrt{1 \pm g^{2}/g_{c}^{2}}$, described by $\hat{c}$ and $\hat{c}^{\dagger}$. A state $\lvert \psi \rangle$ then evolves as
\begin{equation}
    \lvert \psi(t) \rangle
    = \exp\!\left[- i t \,\hat{c}^{\dagger}\hat{c}\,\omega \sqrt{1 \pm g^{2}/g_{c}^{2}}\right] \lvert \psi \rangle ,
\end{equation}
and the quantum Fisher information for $\omega$ takes the form
\begin{equation}
    \mathcal{I}_{\omega}
    = 4 t^{2}\,\Delta^{2}\hat{c}^{\dagger}\hat{c} \,\left(\partial_{\omega}\,\omega \sqrt{1 \pm g^{2}/g_{c}^{2}}\right)^{2},
\end{equation}
where $t$ is the free evolution (interrogation) time of the effective oscillator, and $\Delta^{2}\hat{c}^{\dagger}\hat{c}$ characterizes the initial state (for a coherent state, $\Delta^{2}\hat{c}^{\dagger}\hat{c} = \lvert \alpha \rvert^{2}$). Near criticality,
\begin{equation}\label{eq:critical_enh}
    \left(\partial_{\omega}\,\omega \sqrt{1 - g^{2}/g_{c}^{2}}\right)^{2}
    = \frac{\big(2 - g^{2}/g_{c}^{2}\big)^{2}}{4\big(1 - g^{2}/g_{c}^{2}\big)}
    \;\approx\; 4 \,\langle \hat{a}^{\dagger}\hat{a} \rangle^{2},
\end{equation}
so the scaling is governed by the critical dependence on the microscopic parameters, equivalently by the excitation number of the original mode $\hat{a}$. In practice, however, the vanishing spectral gap entails long transformation or relaxation times; for instance, in driven–dissipative realizations the steady-state relaxation time grows with the oscillator quality factor and inverse frequency~\cite{gardiner2004quantumnoise}, consistent with $T_{-}$.

In the anti-critical regime,
\begin{equation}
     \left(\partial_{\omega}\,\omega \sqrt{1 + g^{2}/g_{c}^{2}}\right)^{2}
     \;\approx\; 4 \,\langle \hat{a}^{\dagger}\hat{a} \rangle^{2},
\end{equation}
so the quantum Fisher information is likewise enhanced proportionally to the excitation number [compare with Eq.~\eqref{eq:critical_enh}], while the relevant relaxation times decrease as the gap opens. This realizes the anti-critical protocol: quantum-enhancement accompanied not by critical slowing down but by accelerated dynamics. Comparing the characteristic timescales (or energy scales), critical [Eq.~\eqref{eq:Tcrit}] versus anti-critical [Eq.~\eqref{eq:anticT}], shows that, for the quantum Rabi model, the anti-critical scheme could attain the same scaling with $\langle \hat{a}^{\dagger}\hat{a} \rangle$ while being faster by a factor proportional to $\langle \hat{a}^{\dagger}\hat{a} \rangle^{2}$ assuming the same level of correlations.

\emph{Conclusion \& Outlook.}---In this work we have shown that in interacting quantum systems the quantum Fisher information does not necessarily need to diverge in order to achieve quantum-enhanced precision. This counter-intuitive result follows from the fact that in critical metrology the relevant timescales grow upon approaching the critical point, whereas, in anti-critical metrology the timescale decreases by exploiting gap opening. Consequently, criticality leads to an enhancement of the quantum Fisher information that need not originate from quantum correlations (see End Matter for a demonstration). By contrast, anti-criticality leads to a suppression of the quantum Fisher information because of the shortening timescale. Nevertheless, comparable quantum enhancement can be achieved for either strategy. However, this viewpoint is often obscured by improper accounting of practically relevant resources, which can be remedied by employing the gap-regularized quantum Fisher information.

To illustrate these principles, we consider the paradigmatic quantum Rabi model, which has been extensively studied in the context of critical metrology and exhibits a finite-component quantum phase transition. By carefully accounting for relevant resources—in particular, the intrinsic timescale set by the energy gap—we find that, for a fixed number of excitations, the anti-critical protocol can achieve the same sensitivity as the conventional critical protocol on significantly shorter characteristic timescales. These results are captured by the operational figure of merit introduced here, the gap-regularized quantum Fisher information. More generally, while critical quantum metrology is limited by long timescales, anti-critical metrology may require strong interactions and tunability of the spectral gap. Nevertheless, the proposed approach highlights that faster dynamics, rather than gap closing, can serve as a resource for quantum metrology. 

As an outlook, our results suggest a new route for quantum metrology in interacting systems by exploiting gap opening and spectral engineering to generate useful quantum correlations. This perspective indicates that non-linear interactions, long-range couplings, or driven–dissipative mechanisms could be harnessed to design sensing protocols in which enhanced precision arises not from critical slowing down but from accelerated dynamics. Finite-size systems appear particularly promising in this context, as the exploitable quantum correlations in anti-critical metrology may ultimately be limited not by the energy gap but by the finite dimensionality of the accessible Hilbert space~\cite{PhysRevA.105.042620,tobe}. By contrast, critical metrology can offer advantages when the available time is sufficiently long to allow for adiabatic state preparation, decoherence is negligible on the relevant timescales, and the number of repetitions $\mu$ can be large. Anti-critical metrology should therefore be viewed as a complementary strategy that excels under realistic constraints of finite coherence and limited interrogation time, where rapid preparation and probing are essential and time itself is a costly resource. 

The optimal strategy is thus context dependent, determined by the spectral gap structure, achievable preparation and interrogation times, and the surrounding noise environment. In practice, candidate protocols can be benchmarked using the gap-regularized quantum Fisher information $\bar{\mathcal{I}}_{\omega} = \delta^2 \mathcal{I}_\omega$ together with experimentally accessible timescales. Overall, the anti-critical paradigm opens a largely unexplored landscape for quantum sensing. Exploring such strategies—and their practical implementations—may considerably broaden the toolbox of quantum-enhanced metrology in interacting many-body systems while clarifying the interplay between quantum correlations and many-body spectral properties.

\emph{Acknowledgments.}---G.M. acknowledges support from Taighde Eireann Research Ireland under grant number 24/EPSRC/4121. Both authors would like to acknowledge fruitful discussions with Paolo Abiuso.

\bibliography{bib.bib}

\newpage
\onecolumngrid 
\subsection{\large{{End Matter}}}
\twocolumngrid


In the main text, the gap-regularized quantum Fisher information \(\bar{\mathcal{I}}_\omega = \delta^2 \mathcal{I}_\omega\) is proposed as a more meaningful figure of merit within the context of ground state quantum metrology. To motivate this proposal, we consider the paradigmatic one-dimensional Ising model:
\begin{equation}
    \hat{H} = \omega \sum_{i=1}^N \hat{\sigma}_z^{(i)} - g \sum_{i=1}^{N}  \hat{\sigma}_x^{(i)} \hat{\sigma}_x^{(i+1)}\;,
\end{equation}
with transverse field \(\omega\) and spin-coupling \(g\). We impose periodic boundary conditions and \(\hat{\sigma}_\alpha^{(i)}\) denotes the \(\alpha\)-component Pauli operator (\(\alpha=x,y,z\)) located at site \(i\). Applying a Jordan-Wigner transformation and going to the momentum space, the Hamiltonian decouples into independent \((k,-k)\) sectors:
\begin{equation}
    \hat{H}_k = 2 \Psi_k^\dagger [(\omega - g \cos{k}) \hat{\sigma}_z + (g \sin{k}) \hat{\sigma}_x]\hat{\Psi}_k\;,
\end{equation}
where \(\hat{\Psi}_k^{\dagger} = (\hat{c}_k^\dagger~~\hat{c}_{-k})\) and the allowed momenta are \( k = \pi(2n+1)/N\) with \(n = 0,1,2...N/2-1\). The eigenstates of each momentum-space block \(\hat{H}_k\) are given by:
\begin{align}
    |\psi_g\rangle_k &= \cos{(\theta_k/2)}|0\rangle_k + \sin{(\theta_k/2)}|1\rangle_k \;, \\
    |\psi_e\rangle_k &= \sin{(\theta_k/2)}|0\rangle_k - \cos{(\theta_k/2)}|1\rangle_k \;, 
\end{align}
with corresponding dispersion relation
\begin{equation}
     \epsilon_k = \pm 2\sqrt{g^2 + \omega^2 - 2 g \omega \cos{k}}\;,
\end{equation}
and the Bogoliubov angle defined as
\begin{equation}
    \tan{\frac{\theta_k}{2}} = \frac{g \sin{k}}{g \cos{k} - \omega}\;.
\end{equation}
The many-body ground state for an \(N\)-site system is the tensor product \(|\psi_{GS}\rangle = \bigotimes_{k>0} |\psi_g\rangle_k\) of the ground state of each momentum space block \(\hat{H}_k\). Given that the quantum Fisher information is additive under a tensor product, one can readily compute the quantum Fisher information with respect to the transverse field using Eq.~\eqref{eq:qfi_gs} to obtain:
\begin{equation}
    \label{eq:tfim_qfi}
    \mathcal{I}_\omega = \sum_k \frac{g^2 \sin^2{k}}{(g^2 + \omega^2 - 2g\omega \cos{k})^2} \equiv \sum_k \frac{16 g^2\sin^2{k}}{(\epsilon_k)^4}\;.
\end{equation}
The summation arises from the contributions of different momentum modes to the total quantum Fisher information. Evaluating Eq.~\eqref{eq:tfim_qfi} at the critical point, \(g = \omega\)
\begin{equation}\label{eq:conqfi}
    \mathcal{I}_\omega = \frac{N^2 - N}{8 \omega^2}\;,
\end{equation}
shows the quantum Fisher information scales quadratically with system size \(N\), i.e., Heisenberg scaling when the particle number is treated as the relevant resource. However, in this form it is difficult to discern the origin of the quantum enhanced sensitivity. To develop an understanding of the underlying mechanism responsible for the Heisenberg scaling, we utilize the spectral representation of the quantum Fisher information: 
\begin{equation}
    \label{eq:qfi_leh}
    \mathcal{I}_\omega = 4 \sum_{n\neq0} \frac{|\langle \psi_n(\omega)| \partial_\omega \hat{H} | \psi_0(\omega)\rangle|^2}{[E_n - E_0 ]^2}\;,
\end{equation}
where the numerator encodes the Hamiltonian sensitivity to infinitesimal variations in \(\omega\) and the denominator captures the sensitivity arising from the spectral gaps and related timescales. Our goal is to disentangle the respective contributions of the spectral gaps and intrinsic many-body sensitivity to the observed Heisenberg scaling. Noting that \(\partial_\omega \hat{H} = \sum_k \partial_\omega \hat{H}_k = 2\sum_k \hat{\Psi}^\dagger\hat{\sigma}_z\hat{\Psi}\), and since \(\partial_\omega \hat{H}\) is a sum of one-block operators, the only excited many-body states that have non-zero overlap with \(\partial_\omega \hat{H}|\psi_{GS}\rangle\) are those in which exactly only one \(k\)-block is promoted from \(|\psi_g\rangle_k\) to \(|\psi_e\rangle_k\):
\begin{equation}
    |\psi_n\rangle \equiv |\psi_k^{(1)}\rangle = |\psi_e\rangle_k \bigotimes_{q\neq k}|\psi_g\rangle_q\;.
\end{equation}
It then follows that the relevant overlaps are given by \(\langle\psi_k^{(1)}|\partial_\omega\hat{H}|\psi_{GS} \rangle = \langle \psi_e | \partial_\omega \hat{H}_k | \psi_g\rangle_k = 2 \sin{\theta_k}\). This is because all other blocks contribute identity overlaps equal to one, and terms with \(q \neq k\) vanish because \(\langle\psi_e|\psi_g\rangle_k = 0\). Therefore, the quantum Fisher information in spectral form can be written as
\begin{equation}
     \mathcal{I}_\omega =  4\sum_{k>0} \mathcal{N}_k \mathcal{D}_k^{-1}\;.
\end{equation}
The numerator appearing in the \(k\)-th contribution is given by:
\begin{align}
    \mathcal{N}_k &= \frac{g^2\sin^2{k}}{g^2 + \omega^2 - 2 g \omega \cos{k} }\;,
\end{align}
and similarly for the denominator
\begin{equation}
    \mathcal{D}_k^{-1} = [4(g^2 + \omega^2 - 2 g \omega \cos{k})]^{-1}\;,
\end{equation}
as the excitation for each contributing block is just \(\Delta E_k = E_{k}^{(1)} - E_{g}\), where $E_k^{(1)}$ is the energy eigenvalue of the $ |\psi_k^{(1)}\rangle$ state. Evaluating both expressions at the critical point, \(g = \omega\), yields:
\begin{equation}
    \mathcal{N}_k = \cos^2{\frac{k}{2}}\;,
\end{equation}
and
\begin{equation}
    \mathcal{D}_k^{-1} = \left[16 \omega^2 \sin^2{\frac{k}{2}}\right]^{-1}\;.
\end{equation}
At the critical point \(g = \omega\), the dominant contribution comes from the smallest momentum mode \(k = \pi/N\). Expanding around this point and dropping higher order terms, the numerator becomes
\begin{equation}
    \mathcal{N}_k \simeq 1 - \frac{1}{4}\left(\frac{\pi}{N}\right)^2\;,
\end{equation}
which approaches a constant in the thermodynamic limit. This indicates that the numerator, which physically quantifies the transition amplitude induced by an infinitesimal change in \(\omega\), does not itself exhibit any quantum enhancement associated with criticality and associated correlations. Consequently, it does not contribute to the emergence of genuine Heisenberg scaling. In contrast, a similar analysis for the denominator at the critical point reveals 
\begin{equation}
    \mathcal{D}_k^{-1} \simeq \frac{N^2}{4 \pi^2 \omega^2}\;,
\end{equation}
i.e., that the apparent Heisenberg scaling in this model is a direct consequence of gap closing at criticality and the associated timescale, not of enhanced matrix elements of many-body coherence in the numerator. This analysis also clarifies the connection to critical slowing down and motivates the use of the gap-regularized quantum Fisher information. The same closure of the spectral gap responsible for the divergence of the quantum Fisher information implies that the system becomes increasingly slow to respond, with characteristic timescale diverging as \(T \sim \delta^{-1} = N^{-1}\) for the Ising model. Consequently, although conventional quantum Fisher information~\eqref{eq:conqfi} suggests enhanced metrological sensitivity, this enhancement is governed entirely by the growing cost in the preparation and interrogation time:
\begin{align}
    \mathcal{I}_\omega \propto N^0T^2\;.
\end{align}
By introducing the gap-regularized quantum Fisher information, one effectively factors out this trivial amplification due to vanishing energy scales. The resulting figure of merit isolates the intrinsic sensitivity of the state to parameter changes, independent of the timescale, and therefore provides a more operationally meaningful measure of metrological usefulness. 

\newpage
\onecolumngrid
\renewcommand{\theequation}{S\arabic{equation}}
\renewcommand{\thefigure}{S\arabic{figure}}
\renewcommand{\thetable}{S\arabic{table}}


\subsection{\large{{Supplemental Material for\\
 ``Mind the Gap: Anti-Critical Quantum Metrology''}}}



\section{Introduction}

This Supplemental Material provides additional derivations, analytical calculations, and numerical details supporting the results presented in the main text.

\section{Quantum Fisher information for a harmonic oscillator ground state}\label{A:QFI_HO}
Here we present the calculation of the quantum Fisher information for the ground state of the quantum Rabi model (mapped to a harmonic oscillator). To this end, we use the following formula
\begin{align}\label{eq:qfiGSsum}
    \mathcal{I}_\omega = 4 \sum_{n \neq 0} \frac{|\langle \psi_n(\omega)| \partial_\omega \hat{H}(\omega) | \psi_0(\omega) \rangle |^2}{\left[E_n(\omega) - E_0(\omega)\right]^2},
\end{align}
expressed through $\omega$-dependent eigenstates $| \psi_n(\omega) \rangle$ and eigenvalues $E_n(\omega)$ of the Hamiltonian $\hat{H}(\omega)$. For the squeezed harmonic oscillator Hamiltonian
\begin{align}
     \hat{H}\approx \omega \hat{a}^\dagger \hat{a}   \pm \frac{g^2}{4\Omega} (\hat{a} + \hat{a}^\dagger)^2,
\end{align}
where $\pm$ indicates closing or opening the energy gap, it is straightforward to show that only one term from the sum over $n$ survives, leading to 
\begin{align}
\begin{split}
    \mathcal{I}_\omega =    \frac{|\langle \psi_2(\omega)|  \hat a^\dagger \hat a  | \psi_0(\omega)\rangle |^2}{ \omega^2\left(1\pm g^2/g_c^2\right)} = \frac{2\langle\hat a^\dagger \hat a\rangle\left(\langle \hat a^\dagger \hat a\rangle +1\right)}{\omega^2(1\pm g^2/g_c^2)}=\frac{\Delta^2 \hat a^\dagger \hat a }{\omega^2 \left(1 \pm g^2/g_c^2\right)}\sim \langle \hat a^\dagger \hat a \rangle^2 T_\pm^2,
 \end{split}
\end{align}
where we have used the fact that $|\psi_n(\omega\rangle  =\exp[\tfrac{\xi_\pm}2(\hat a^{\dagger2}-\hat a^2)]|n\rangle$ and $T_\pm^{-2} \sim  \omega^2(1\pm g^2/g_c^2)$ is related to the energy gap inverse squared. In both cases the quantum Fisher information exhibits quadratic (Heisenberg) scaling with the number of excitations $\langle \hat a^\dagger \hat a\rangle$ and time $T_\pm$ which is related to the energy gap.

Since there is a mapping from the Lipkin-Meshkov-Glick model to a harmonic oscillator related to the Holstein-Primakoff transformation, the quantum Fisher information for the former can be expressed as
\begin{align}
    \mathcal{I}_\omega \approx \Delta^2 \hat S_z \, \delta^{-2},
\end{align}
where $\delta=\omega\sqrt{1-g/\omega}$, which can be seen in Fig.~\ref{fig:fig2}c.

\section{Quantum Fisher information for a squeezed vacuum}\label{A:QFI_SV}
For the sake of completeness, we show the quantum Fisher information for a squeezed vacuum $|\psi_0\rangle$ where the information about the unknown parameter $\omega$ is being imprinted through $\hat H = \omega \hat a^\dagger \hat a$ for time $t$. As a consequence the quantum Fisher information becomes
\begin{align}
    \mathcal{I}_\omega = 4 t^2 \Delta^2 \hat a^\dagger \hat a = 4t^2\times2\langle\hat a^\dagger \hat a\rangle\left(\langle \hat a^\dagger \hat a\rangle +1\right),
\end{align}
where $t$ is the free evolution time and is not related to the energy gap. This expression highlights two key points. First, the quantum Fisher information is directly enhanced by the number fluctuations of the squeezed state, its orientation does not matter. Second, it grows quadratically with the evolution time for a coherent process. 

\section{Quantum Fisher information for an adiabatic time evolution}
\label{A:adiabaticQFI}

We consider a family of Hamiltonians
\begin{equation}
    \hat H(\omega,t)\,|\psi_n(t)\rangle
    =
    E_n(\omega,t)\,|\psi_n(t)\rangle,
\end{equation}
depending smoothly on the parameter $\omega$.
The time evolution is governed by the time-ordered unitary
\begin{equation}
    \hat U(\omega,T)
    =
    \mathcal{T}
    \exp\!\left[-i\!\int_0^T \hat H(\omega,t)\,dt\right].
\end{equation}
The generator of infinitesimal translations in $\omega$ is defined as
\begin{equation}
    \hat G(\omega,T)
    =
    i\,\hat U^\dagger(\omega,T)\,\partial_\omega \hat U(\omega,T).
\end{equation}

\subsection*{Exact integral representation of the generator}

Differentiating the time-ordered exponential and using
$\hat U(t,0)=\mathcal{T}e^{-i\int_0^t\hat H(\omega,\tau)\,d\tau}$,
one obtains the exact identity
\begin{equation}
    \hat G(\omega,T)
    =
    \int_0^T
    \hat U^\dagger(t,0)\,
    \partial_\omega \hat H(\omega,t)\,
    \hat U(t,0)\,dt .
    \label{eq:G-exact}
\end{equation}
Introducing the Heisenberg-picture operator
\begin{equation}
    \hat A_H(t)
    \equiv
    \hat U^\dagger(t,0)\,
    \partial_\omega \hat H(\omega,t)\,
    \hat U(t,0),
\end{equation}
the generator can be written compactly as
\begin{equation}
    \hat G
    =
    \int_0^T \hat A_H(t)\,dt .
\end{equation}

\subsection*{Expectation value of the generator}

We assume that the system is initially prepared in the instantaneous ground state
$|\psi_0(0)\rangle$.
The expectation value of the generator is then
\begin{equation}
    \langle \hat G\rangle
    =
    \int_0^T
    \langle \psi_0(0)|\hat A_H(t)|\psi_0(0)\rangle\,dt.
\end{equation}
Within the adiabatic approximation,
\begin{equation}
    \hat U(t,0)|\psi_n(0)\rangle
    \approx
    e^{-i\theta_n(t)}\,|\psi_n(t)\rangle,
\end{equation}
where
\begin{equation}
    \theta_n(t)
    =
    \int_0^t E_n(\tau)\,d\tau
    -
    i\int_0^t
    \langle \psi_n(\tau)|\partial_\tau\psi_n(\tau)\rangle\,d\tau
\end{equation}
is the sum of the dynamical and Berry phases.
This yields
\begin{equation}
    \langle \hat G\rangle
    \approx
    \int_0^T
    \langle \psi_0(t)|\partial_\omega \hat H(\omega,t)|\psi_0(t)\rangle\,dt ,
\end{equation}
which coincides with the Feynman--Hellmann expression integrated along the adiabatic path.

\subsection*{Second moment of the generator}

The second moment of the generator is given by
\begin{equation}
    \langle \hat G^2\rangle
    =
    \int_0^T\!\!\int_0^T
    \langle \psi_0(0)|
    \hat A_H(t)\,\hat A_H(t')
    |\psi_0(0)\rangle
    \,dt\,dt'.
\end{equation}
Inserting the resolution of the identity
\begin{equation}
    \hat{\mathbb I}
    =
    \sum_n |\psi_n(0)\rangle\langle \psi_n(0)|,
\end{equation}
we obtain
\begin{align}
    \langle \hat G^2\rangle
    &=
    \sum_n
    \int_0^T\!dt
    \int_0^T\!dt'\,
    \langle \psi_0(0)|\hat A_H(t)|\psi_n(0)\rangle
    \langle \psi_n(0)|\hat A_H(t')|\psi_0(0)\rangle .
\end{align}
Using the adiabatic approximation for the matrix elements,
\begin{align}
    \langle \psi_0(0)|\hat A_H(t)|\psi_n(0)\rangle
    &\approx
    e^{\,i[\theta_0(t)-\theta_n(t)]}
    \langle \psi_0(t)|\partial_\omega \hat H(\omega,t)|\psi_n(t)\rangle ,
\end{align}
the double time integral factorizes, yielding
\begin{equation}
    \langle \hat G^2\rangle
    \approx
    \sum_n
    \left|
    \int_0^T\!
    e^{\,i[\theta_0(t)-\theta_n(t)]}
    \langle \psi_0(t)|\partial_\omega \hat H(\omega,t)|\psi_n(t)\rangle
    \,dt
    \right|^2 .
    \label{eq:G2-adiabatic}
\end{equation}
Here we retain only the leading adiabatic contribution and neglect non-adiabatic corrections, which are suppressed by inverse powers of the instantaneous energy gap.

\subsection*{Variance and quantum Fisher information}

The variance of the generator,
\begin{equation}
    \Delta^2 \hat G
    =
    \langle \hat G^2\rangle
    -
    \langle \hat G\rangle^2,
\end{equation}
determines the quantum Fisher information via
$\mathcal I_\omega = 4\,\Delta^2\hat G$.
Combining the results above, the diagonal contribution ($n=0$) cancels identically, yielding
\begin{equation}
    \Delta^2 \hat G
    \approx
    \sum_{n\neq 0}
    \left|
    \int_0^T\!
    e^{\,i[\theta_0(t)-\theta_n(t)]}
    \langle \psi_0(t)|\partial_\omega \hat H(\omega,t)|\psi_n(t)\rangle
    \,dt
    \right|^2,
\end{equation}
where $T$ is the length of the adiabatic time evolution related to the gap inverse [compare with Eq.~\eqref{eq:qfiGSsum}]. The orientation of squeezing in the ground state does not matter as rotating the squeezed vacuum with $\hat a^\dagger \hat a$ for a harmonic oscillator does not change the expectation value of $\hat a^\dagger \hat a$ and similar for spin-squeezed states. In critical metrology $T$ is very large due to closing energy gap which increases the quantum Fisher information, while in anti-critical metrology $T$ is very small due to opening of the energy gap which reduces the quantum Fisher information.

No matter how the adiabatic ramp is optimized, the quantum Fisher information associated with an adiabatic evolution connecting neighboring ground states is given by
\begin{align}
\mathcal{I}_\omega 
= 4 \sum_{n \neq 0} 
\frac{|\langle \psi_n(g)| \partial_\omega \hat{H} | \psi_0(g) \rangle |^2}
{\left[E_n(g) - E_0(g)\right]^2},
\end{align}
and depends only on the end point.
To show this, we rewrite the time integral in terms of the control parameter $g$, using $dt = dg/\dot g$. This yields
\begin{align}
\Delta^2 \hat G
\approx
\sum_{n\neq 0}
\left|
\int_{g_0}^{g_f}
e^{\,i[\theta_0(g)-\theta_n(g)]}
\langle \psi_0(g)|\partial_\omega \hat H(\omega,g)|\psi_n(g)\rangle
\,\frac{dg}{\dot g}
\right|^2,
\end{align}
where $\dot g \equiv dg/dt$. Using
\begin{align}
e^{i\theta} = \frac{1}{i\,\partial_g \theta}\,\partial_g e^{i\theta},
\end{align}
we integrate by parts and obtain
\begin{align}
\Delta^2 \hat G
\approx
\sum_{n\neq 0}
\Bigg|
\frac{e^{\,i[\theta_0-\theta_n]}}{i}\,
\frac{\langle \psi_0|\partial_\omega \hat H|\psi_n\rangle}{E_0 - E_n}
\Bigg|_{g_0}^{g_f}
-
\int_{g_0}^{g_f}
\frac{e^{\,i[\theta_0-\theta_n]}}{i}\,
\partial_g
\left[
\frac{\langle \psi_0|\partial_\omega \hat H|\psi_n\rangle}{E_0 - E_n}
\right]
dg
\Bigg|^2,
\end{align}
where we used
\begin{align}
\partial_g(\theta_0 - \theta_n) = \frac{E_0(g) - E_n(g)}{\dot g}.
\end{align}

In the adiabatic limit, the second term is suppressed. Indeed, the phase factor $e^{i[\theta_0(g)-\theta_n(g)]}$ becomes rapidly oscillatory as $\dot g \to 0$, while the remaining prefactor varies smoothly with $g$. By the Riemann--Lebesgue lemma (or, equivalently, by repeated integration by parts), such integrals vanish in the adiabatic limit. Consequently, only the boundary contribution survives, yielding
\begin{align}
\Delta^2 \hat G
\approx
\sum_{n\neq 0}
\left|
\frac{\langle \psi_0(g)|\partial_\omega \hat H(\omega,g)|\psi_n(g)\rangle}
{E_0(g)-E_n(g)}
\Bigg|_{g_0}^{g_f}
\right|^2.
\end{align}

For $g_0=0$, the initial contribution vanishes, and we obtain
\begin{align}
\Delta^2 \hat G
\approx
\sum_{n\neq 0}
\frac{|\langle \psi_n(g_f)|\partial_\omega \hat H(\omega,g_f)|\psi_0(g_f)\rangle|^2}
{\left[E_n(g_f)-E_0(g_f)\right]^2},
\end{align}
which coincides with the static quantum Fisher information. More generally, the result is independent of the initial point $g_0$. A finite initial value simply corresponds to information already encoded in the ground state prior to the considered evolution.

\begin{figure}[htb!]
    \centering
    \includegraphics[width=0.7\linewidth]{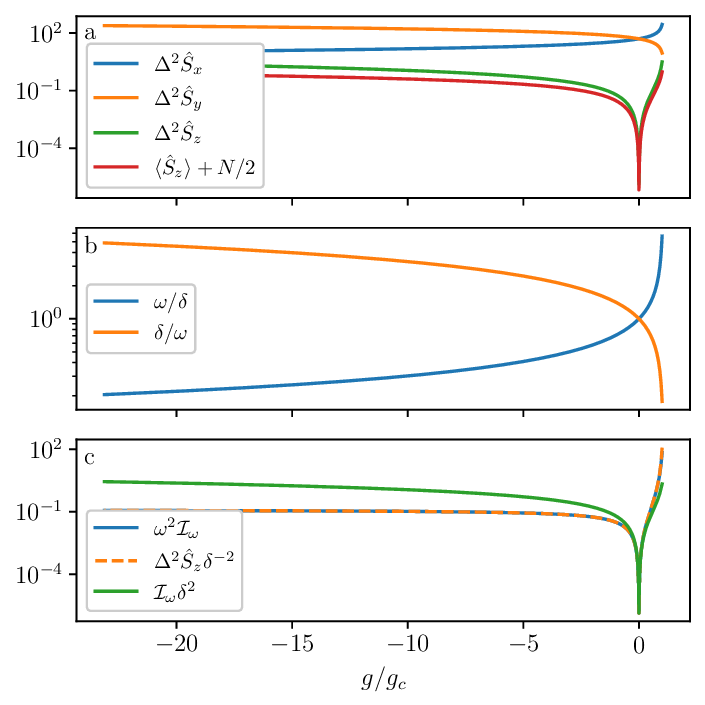}
    \caption{Lipkin-Meshkov-Glick model. (a) Variances of collective spin components as a function of $g/g_c$ ($g_c\equiv \omega$). The variances of $\hat S_x$ (blue line) and $\hat S_y$ (orange line) are flipped with respect to each other which shows that quantum correlations can be generated close and away from the critical point. The variance of $\hat S_z$ (green line) grows both close and away from the critical point similarly as its expectation value (red line shifted by $N/2$ for the matter of presentation). (b) The associated energy gap (blue line) and its inverse (orange line) are very closely related to variances of $\hat S_x$ and $\hat S_y$ which points at a close connection between these quantities. Finally, in (c), we show the quantum Fisher information (blue line) as a function of $g/g_c$ which is related to the expectation value of $\hat S_z$ weighted by the squared inverse of the energy gap (dashed orange line). Naively it seems that the quantum Fisher information is superior close to the critical point, however, this is caused by the fact that the characteristic time scale artificially elevates the quantum Fisher information. Once the characteristic time is factored out (green line), it turns out that quantum Fisher information can be equally large far away from the critical point and achieved much faster due to decreasing energy gap. } 
    \label{fig:fig2}
\end{figure}

\section{Lipkin-Meshkov-Glick model}\label{A:LMG}
In the main text, we have focused on a pedagogical example of the quantum Rabi model, which can be understood from the viewpoint of an elementary harmonic oscillator. Here, we elucidate the principle of anti-critical metrology using a many-body quantum system of interacting identical and indistinguishable spins. Specifically, we consider a symmetric all-to-all interacting Lipkin-Meshkov-Glick model
\begin{align}
    \hat H = \omega \hat S_z -\frac{g}{N}\hat S_x^2,
\end{align}
where $\hat S_\alpha = \tfrac12\sum_{i=1}^N \hat \sigma_\alpha$ with $\alpha = x,y,z$ are the collective spin operators, $\omega$ is the natural spin frequency, $g$ is the interaction strength, and $N$ is the number of spins. Importantly, we restrict our analysis to the symmetric subspace of the Hilbert space, i.e., states with maximal total spin $S=N/2$, which are fully symmetric under particle exchange and capture the collective behavior relevant for critical and anti-critical metrology.  

In principle, the Lipkin-Meshkov-Glick model can be considered in the thermodynamic limit, where there exists a mapping to a harmonic oscillator. Here, however, we deliberately perform numerical calculations on $N=200$ spins to show that a system does not have to map to a harmonic oscillator to benefit from the anti-critical metrology approach. The behavior of the energy gap, the variances of collective spin operators (related to spin squeezing), and the quantum Fisher information both near and away from the critical point are shown in Fig.~\ref{fig:fig2}. The variances of the collective spin components in the ground state [panel (a)] closely follow the energy gap between the ground and first excited states, $\delta = E_1 - E_0$ [panel (b)]. Panel (c) illustrates that the quantum Fisher information (blue line) is essentially the variance of $\hat{S}_z$ weighted by the inverse square of the energy gap (orange dashed line). The enhancement of the quantum Fisher information near the critical point therefore arises primarily from the closing of the gap, which lengthens the characteristic time scale of the dynamics. Conversely, away from criticality, the gap increases, reducing the relevant time scale and leading to a smaller, approximately constant quantum Fisher information despite increasing $\Delta^2 \hat{S}_z$. When the quantum Fisher information is normalized by this time scale (green line), it simply recovers the spin variance $\Delta^2 \hat{S}_z$, which can remain large even far from the critical point [green line in Fig.~\ref{fig:fig2}a].

\section{Transverse field Ising model}\label{A:TFIM}
Now let us turn to another paradigmatic model exhibiting a quantum phase transition, namely, the transverse-field Ising model with nearest-neighbor interactions. The Hamiltonian of the model reads
\begin{align}
    \hat H = \omega \sum_{i=1}^N \hat \sigma_z^{(i)} 
    - g \sum_{i=1}^N \hat \sigma_x^{(i)} \hat \sigma_x^{(i+1)},
\end{align}
where \(g\) denotes the strength of the spin--spin interaction, and we have assumed periodic boundary conditions, 
\(\hat \sigma_x^{(N+1)} \equiv \hat \sigma_x^{(1)}\). 
The system undergoes a quantum phase transition at the critical point \(g=\omega \), 
separating the paramagnetic and ferromagnetic phases. 

\begin{figure}[htb!]
    \centering
    \includegraphics[width=0.7\linewidth]{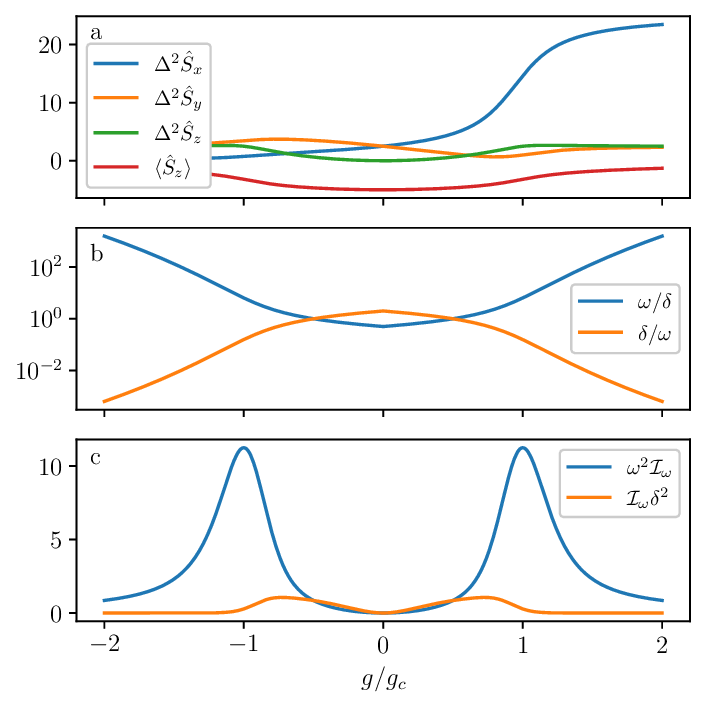}
    \caption{Transverse field Ising model. (a) Variances of $\hat S_x$ (blue line), $\hat S_y$ (orange line), and $\hat S_z$ (green line) as well as expectation value of $\hat S_z$ (red line) as a function of $g/g_c$). As can be seen, correlations can be equally large close and away from the critical point $g/g_c=1$. However the energy gap (b) is symmetric with respect to $g/g_c=0$ point so the gap never increases with increasing coupling. In (c), we see the effect on the quantum Fisher information (blue line) which is symmetric because the correlations depend on $|g/g_c|$, and once divided by the energy gap (orange line) is also symmetric because the gap also depends on $|g/g_c|$ and not on its sign. In this case, anti-critical metrology approach cannot work. In the simulations we have set $N=10$.}
    \label{fig:fig4}
\end{figure}

For the anti-critical metrology protocol to be effective, 
many-body interactions must open a finite energy gap that stabilizes the correlated phase away from criticality. However, as illustrated in Fig.~\ref{fig:fig4}, this mechanism is not universal. In the case of the transverse field Ising model, the energy gap inevitably closes regardless of the interaction strength [see Fig.~\ref{fig:fig4}b] similar as in the quantum Rabi model from the main text if the spin is in its ground state. Consequently, this model does not allow for an anti-critical regime where strong correlations coexist with a increased excitation gap [see Fig.~\ref{fig:fig4}c].

\section{Transverse field Ising model with transverse interactions}\label{A:TI}
\begin{figure}[htb!]
    \centering
    \includegraphics[width=0.7\linewidth]{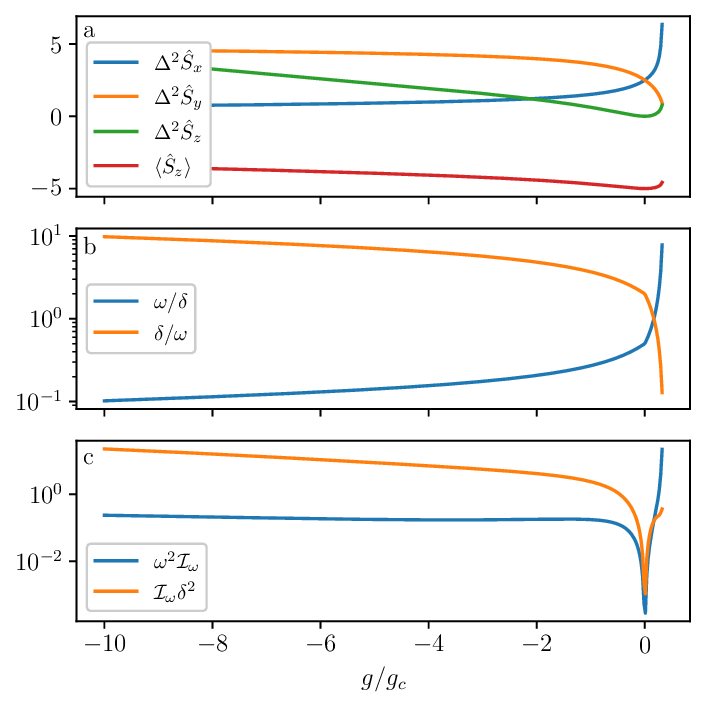}
    \caption{Transverse field Ising model with transverse interactions. (a) Variances of $\hat S_x$ (blue line), $\hat S_y$ (orange line), and $\hat S_z$ (green line) as well as expectation value of $\hat S_z$ (red line) as a function of $g/g_c$). As can be seen, again correlations can be equally large close and away from the critical point $g/g_c=1$. However, the energy gap (b) is asymmetric with respect to $g/g_c=0$ point so the gap can increase with the increasing coupling. In (c), we see the effect on the quantum Fisher information (blue line) which is now asymmetric because the correlations depend on $g/g_c$ in a different way than the gap. In this case, anti-critical metrology approach can work as evidenced by the quantum Fisher information divided by the energy gap (orange line). In the simulations we have set $N=10$.}
    \label{fig:fig5}
\end{figure}

The lack of an energy gap opening can be fixed by adding extra nearest-neighbor interactions in the transverse directions. In this case, the Hamiltonian becomes
\begin{align}
    \hat H = \omega \sum_{i=1}^N \hat \sigma_z^{(i)} 
    - g \sum_{i=1}^N \left(\hat \sigma_x^{(i)} \hat \sigma_x^{(i+1)} - \hat \sigma_z^{(i)} \hat \sigma_z^{(i+1)}\right),
\end{align}
which is a special variant of the anisotropic XYZ model, where the interactions along $y$ direction are absent. Adding transverse interactions is not the only way to increase the energy gap, but here we only want to demonstrate that neither all-to-all interactions nor high symmetry of the eigenstates are necessary to harness the anti-critical metrology framework. In particular, unlike in the Lipkin-Meshkov-Glick model, it is not necessary to restrict the dynamics to the maximally symmetric subspace. The results of numerical simulations for $N=10$ spins are presented in Fig.~\ref{fig:fig5}. In panel (a), we show the build-up of correlations as a function of $g/g_c$. In panel (b), we show the energy gap and its inverse, which exhibit the desired behavior of growing with respect to the non-interacting case $g=0$. In panel (c), we show the quantum Fisher information, which is maximal close to the critical point (blue line); however, once divided by the characteristic timescale related to the inverse of the energy gap, we see that the quantum Fisher information can be more optimal away from the critical point where the gap increases (orange line). The results are very similar to the results obtained for the Lipkin-Meshkov-Glick model (see Fig.~\ref{fig:fig2}) which confirms that neither long-range interactions nor additional constrains on the symmetry of the eigenstates are required for the anti-critical metrology framework to work efficiently.

\section{Transverse-field Ising model with a tilted field}
\label{A:TFIM-tilted}

An alternative route to generating effective transverse interactions---without explicitly adding extra interaction terms---is to supplement the transverse-field Ising model with a {tilted} field. Unlike in the previous Section, where transverse interactions arise from explicit $\hat \sigma_z^{(i)}\hat \sigma_z^{(i+1)}$ couplings, here the mechanism is indirect. Tilting the field away from the $z$-axis breaks the $\mathbb{Z}_2$ symmetry and, in combination with nearest--neighbour $x$-Ising coupling, induces effective interactions that mix longitudinal and transverse spin components.

We consider the Ising Hamiltonian with a nearest-neighbour $x$-Ising interaction and a field of fixed magnitude~$\omega$ pointing in the $(x,z)$-plane:
\begin{align}
    \hat H
    = \omega\cos\phi\sum_{i=1}^N \hat\sigma_z^{(i)}
    + \omega\sin\phi\sum_{i=1}^N \hat\sigma_x^{(i)}
    - g\sum_{i=1}^N \hat\sigma_x^{(i)}\hat\sigma_x^{(i+1)} .
    \label{eq:ising-tilted}
\end{align}
For $\phi=0$ one recovers the standard transverse-field Ising model. Deviating from $\phi=0$ lifts the $\mathbb{Z}_2$ symmetry and modifies the low-energy structure of the model, often enhancing the gap depending on $g/\omega$ and $\phi$.

\begin{figure}[htb!]
    \centering
    \includegraphics[width=0.7\linewidth]{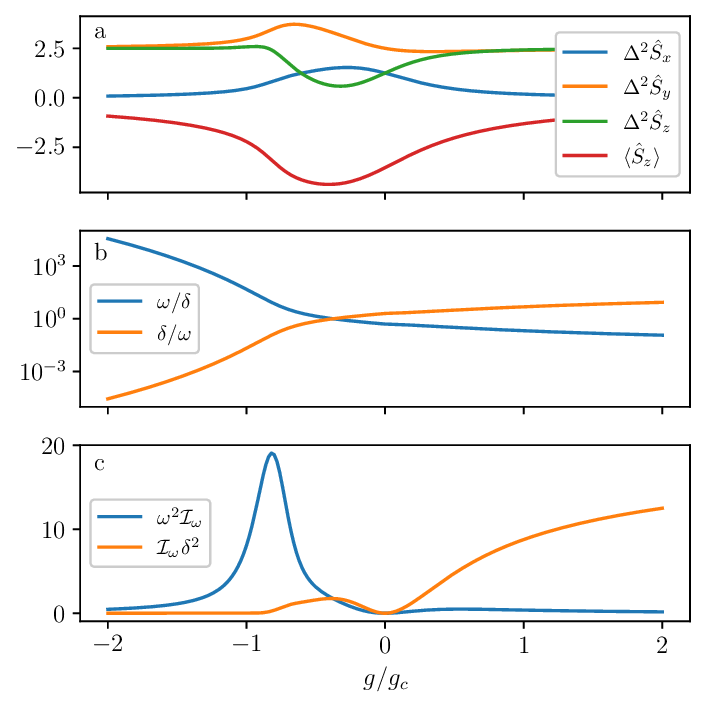}
    \caption{Transverse field Ising model with a tilted field $\phi=\pi/2$. (a) Variances of $\hat S_x$ (blue line), $\hat S_y$ (orange line), and $\hat S_z$ (green line) as well as expectation value of $\hat S_z$ (red line) as a function of $g/g_c$). In interacting quantum systems the correlations can be equally large for positive and negative couplings $g/g_c$, however, the energy gap (b) can be asymmetric with respect to $g/g_c=0$ point so the gap can increase with the increasing coupling. In (c), we see the effect on the quantum Fisher information (blue line) which is now asymmetric because the correlations depend on $g/g_c$ in a different way than the gap. In this case, anti-critical metrology approach can work as evidenced by the quantum Fisher information divided by the energy gap (orange line). In the simulations we have set $N=10$.}
    \label{fig:fig6}
\end{figure}

It is convenient to rotate all spins such that the field becomes aligned with the new $z'$ direction. We introduce a global rotation around the $y$-axis,
\begin{align}
    \hat R(\phi) = \prod_{i=1}^N e^{-i\frac{\phi}{2}\hat\sigma_y^{(i)}} .
\end{align}
The rotated Pauli operators,
\begin{align}
    \hat\sigma_{\alpha'}^{(i)}
    = \hat R(\phi)\,\hat\sigma_\alpha^{(i)}\,\hat R^\dagger(\phi),
\end{align}
satisfy the identities
\begin{align}
    \hat\sigma_z &= \cos\phi\,\hat\sigma_{z'} + \sin\phi\,\hat\sigma_{x'}, \\
    \hat\sigma_x &= \cos\phi\,\hat\sigma_{x'} - \sin\phi\,\hat\sigma_{z'} .
\end{align}
Using these relations, the field term in Eq.~\eqref{eq:ising-tilted} becomes diagonal:
\begin{align}
    \omega\cos\phi\,\hat\sigma_z
    + \omega\sin\phi\,\hat\sigma_x
    = \omega\,\hat\sigma_{z'} .
\end{align}
Thus, the rotation simply maps the tilted field to a standard transverse field of magnitude~$\omega$.

The nearest-neighbour interaction transforms as
\begin{align}
\begin{split}
    \hat\sigma_x^{(i)}\hat\sigma_x^{(i+1)}
    =&\;
      \cos^2\phi\,
      \hat\sigma_{x'}^{(i)}\hat\sigma_{x'}^{(i+1)}
    \;-\;
      \cos\phi\sin\phi\,
      \left(
          \hat\sigma_{x'}^{(i)}\hat\sigma_{z'}^{(i+1)}
        + \hat\sigma_{z'}^{(i)}\hat\sigma_{x'}^{(i+1)}
      \right)
    \;+\;
      \sin^2\phi\,
      \hat\sigma_{z'}^{(i)}\hat\sigma_{z'}^{(i+1)},
\end{split}
\end{align}
thus, the rotated Hamiltonian $\hat H'=\hat R(\phi)\hat H\hat R^\dagger(\phi)$ becomes
\begin{align}
\begin{split}
    \hat H'
    =& -g\sum_{i=1}^N
    \Big[
        \cos^2\phi\;\hat\sigma_{x'}^{(i)}\hat\sigma_{x'}^{(i+1)}
       -\cos\phi\sin\phi\,
        \left(
            \hat\sigma_{x'}^{(i)}\hat\sigma_{z'}^{(i+1)}
          + \hat\sigma_{z'}^{(i)}\hat\sigma_{x'}^{(i+1)}
        \right)
        +\sin^2\phi\;\hat\sigma_{z'}^{(i)}\hat\sigma_{z'}^{(i+1)}
    \Big]
    \;-\;
    \omega\sum_{i=1}^N \hat\sigma_{z'}^{(i)} .
\end{split}
\label{eq:rotated-hamiltonian-tilted}
\end{align}

Equation~\eqref{eq:rotated-hamiltonian-tilted} demonstrates that a simple tilt of the external field generates new interaction channels absent in the original Hamiltonian~\eqref{eq:ising-tilted}, including mixed longitudinal–transverse couplings and an induced $z'z'$ interaction. Thus, while the bare model features only $xx$-type interactions, the rotated picture reveals a richer structure of effective couplings. These additional correlations can enhance the excitation gap in a manner analogous to the explicit transverse interactions discussed in the main text. From the perspective of anti-critical metrology, this highlights that neither fine-tuned symmetries nor all-to-all interactions are necessary to stabilize the ground state while preserving its enhanced sensitivity. Numerical results supporting these observations are presented in Fig.~\ref{fig:fig6}. For negative couplings, the system behaves like a transverse-field Ising model, exhibiting critical slowing down due to the closing energy gap. In contrast, for positive couplings, the gap and the relevant correlations increase, enabling the anti-critical metrology protocol.

\end{document}